\documentclass[aps,prc,twocolumn,superscriptaddress]{revtex4-2} 

\usepackage{graphicx}
\usepackage{subfigure}
\usepackage{multirow}
\usepackage{hyperref}

\hypersetup{
    colorlinks=true,
    citecolor=blue,
    filecolor=black,
    linkcolor=blue,
    urlcolor=blue,
    pdfauthor=author,
    }

\begin{document}

\title{\texorpdfstring{\( \beta \)}{Beta}-delayed neutron spectroscopy of \texorpdfstring{\( ^{85, 86} \)}{85,86}As with MONSTER at IGISOL}

\author{A.~Pérez de Rada Fiol}
\email[A.~Pérez de Rada Fiol. Corresponding author at CIEMAT, Avenida Complutense 40, 28040 Madrid, Spain. Tel: +34 913466615. Email: ]{alberto.rada@ciemat.es}
\author{T.~Martínez}
\author{D.~Cano-Ott}
\affiliation{Centro de Investigaciones Energéticas, Medioambientales y Tecnológicas (CIEMAT), 28040 Madrid, Spain}
\author{H.~Penttilä}
\affiliation{University of Jyväskylä, Department of Physics, Accelerator Laboratory, P.O. Box 35 (JYFL), 40014 University of Jyväskylä, Finland}

\author{J.~Agramunt}
\affiliation{Instituto de Física Corpuscular (IFIC), CSIC-Universidad de Valencia, 46071 Valencia, Spain}
\author{V.~Alcayne}
\affiliation{Centro de Investigaciones Energéticas, Medioambientales y Tecnológicas (CIEMAT), 28040 Madrid, Spain}
\author{A.~Algora}
\affiliation{Instituto de Física Corpuscular (IFIC), CSIC-Universidad de Valencia, 46071 Valencia, Spain}
\author{S.~Alhomaidhi}
\affiliation{Institut für Kernphysik, Technische Universität Darmstadt, 64289 Darmstadt, Germany}
\affiliation{GSI Helmholtzzentrum für Schwerionenforschung, 64291 Darmstadt, Germany}
\author{J.~Äystö}
\affiliation{University of Jyväskylä, Department of Physics, Accelerator Laboratory, P.O. Box 35 (JYFL), 40014 University of Jyväskylä, Finland}
\affiliation{Helsinki Institute of Physics (HIP), 00014 University of Helsinki, Finland}
\author{K.~Banerjee}
\affiliation{Variable Energy Cyclotron Centre (VECC), 700064 Kolkata, India}
\affiliation{Homi Bhabha National Institute (HBNI), 400094 Anushakti Nagar, Mumbai, India}
\author{O.~Beliuskina}
\affiliation{University of Jyväskylä, Department of Physics, Accelerator Laboratory, P.O. Box 35 (JYFL), 40014 University of Jyväskylä, Finland}
\author{J.~Benito}
\affiliation{Universidad Complutense de Madrid (UCM), Grupo de Física Nuclear, 28040 Madrid, Spain}
\author{C.~Bhattacharya}
\affiliation{Variable Energy Cyclotron Centre (VECC), 700064 Kolkata, India}
\affiliation{Homi Bhabha National Institute (HBNI), 400094 Anushakti Nagar, Mumbai, India}
\author{F.~Calviño}
\author{G.~Cortés}
\affiliation{Universitat Politècnica de Catalunya (UPC), Departament de Física, 08034 Barcelona, Spain}
\author{R.~P.~de Groote}
\author{A.~de Roubin}
\affiliation{University of Jyväskylä, Department of Physics, Accelerator Laboratory, P.O. Box 35 (JYFL), 40014 University of Jyväskylä, Finland}
\author{C.~Delafosse}
\affiliation{IJCLab, Université Paris-Saclay, CNRS/IN2P3, 91405 Orsay Cedex, France}
\affiliation{University of Jyväskylä, Department of Physics, Accelerator Laboratory, P.O. Box 35 (JYFL), 40014 University of Jyväskylä, Finland}
\author{C.~Domingo-Pardo}
\affiliation{Instituto de Física Corpuscular (IFIC), CSIC-Universidad de Valencia, 46071 Valencia, Spain}
\author{S.~Geldhof}
\author{W.~Gins}
\author{M.~Hukkanen}
\author{A.~Jokinen}
\author{A.~Kankainen}
\affiliation{University of Jyväskylä, Department of Physics, Accelerator Laboratory, P.O. Box 35 (JYFL), 40014 University of Jyväskylä, Finland}
\author{J.~Lerendegui-Marco}
\affiliation{Instituto de Física Corpuscular (IFIC), CSIC-Universidad de Valencia, 46071 Valencia, Spain}
\affiliation{Universidad de Sevilla (US), Facultad de Física, 41012 Sevilla, Spain}
\author{J.~Llanes Gamonoso}
\affiliation{Centro de Investigaciones Energéticas, Medioambientales y Tecnológicas (CIEMAT), 28040 Madrid, Spain}
\author{I.~Matea}
\affiliation{IJCLab, Université Paris-Saclay, CNRS/IN2P3, 91405 Orsay Cedex, France}
\author{E.~Mendoza}
\affiliation{Centro de Investigaciones Energéticas, Medioambientales y Tecnológicas (CIEMAT), 28040 Madrid, Spain}
\author{A.~K.~Mistry}
\affiliation{Institut für Kernphysik, Technische Universität Darmstadt, 64289 Darmstadt, Germany}
\affiliation{GSI Helmholtzzentrum für Schwerionenforschung, 64291 Darmstadt, Germany}
\author{D.~A.~Nesterenko}
\affiliation{University of Jyväskylä, Department of Physics, Accelerator Laboratory, P.O. Box 35 (JYFL), 40014 University of Jyväskylä, Finland}
\author{J.~Plaza}
\affiliation{Centro de Investigaciones Energéticas, Medioambientales y Tecnológicas (CIEMAT), 28040 Madrid, Spain}
\author{I.~Pohjalainen}
\author{S.~Rinta-Antila}
\affiliation{University of Jyväskylä, Department of Physics, Accelerator Laboratory, P.O. Box 35 (JYFL), 40014 University of Jyväskylä, Finland}
\author{P.~Roy}
\affiliation{Variable Energy Cyclotron Centre (VECC), 700064 Kolkata, India}
\affiliation{Homi Bhabha National Institute (HBNI), 400094 Anushakti Nagar, Mumbai, India}
\author{A.~Sanchez-Caballero}
\affiliation{Centro de Investigaciones Energéticas, Medioambientales y Tecnológicas (CIEMAT), 28040 Madrid, Spain}
\author{J.~L.~Taín}
\affiliation{Instituto de Física Corpuscular (IFIC), CSIC-Universidad de Valencia, 46071 Valencia, Spain}
\author{D.~Villamarín}
\affiliation{Centro de Investigaciones Energéticas, Medioambientales y Tecnológicas (CIEMAT), 28040 Madrid, Spain}
\author{M.~Vilen}
\affiliation{University of Jyväskylä, Department of Physics, Accelerator Laboratory, P.O. Box 35 (JYFL), 40014 University of Jyväskylä, Finland}

\date{\today}

\begin{abstract}
  The \( \beta \)-delayed neutron emission in the \( ^{85, 86} \)As \( \beta \)-decays has been measured at the Ion Guide Isotope Separator On Line facility of the Accelerator Laboratory of the University of Jyväskylä. The complete \( \beta \)-decays have been studied with a complex setup that consists of a plastic scintillator for \( \beta \)-particles, MONSTER---the MOdular Neutron time-of-flight SpectromeTER---for neutrons, and a high-purity germanium and four LaBr\( _3 \) crystals for \( \gamma \)-rays. The \( \beta \)-delayed neutron energy distributions have been determined by unfolding the time-of-flight spectra with an innovative methodology based on the iterative Bayesian unfolding method and accurate Monte Carlo simulations. The results obtained for \( ^{85} \)As are in excellent agreement with the existing evaluated data, validating the proposed methodology. In the case of \( ^{86} \)As, a stronger neutron intensity at higher energies than previously predicted is discovered.
\end{abstract}

\maketitle

\section{Introduction}\label{intro}

\( \beta \)-delayed neutron emission was discovered in 1939 by Roberts et al.~\cite{Roberts1939_1, Roberts1939_2}, shortly after the discovery of fission by Meitner et al.\ in 1938~\cite{Meitner1938}. \( \beta \)-delayed two- and three-neutron emission was discovered \( 40 \) years later in \( ^{11} \)Li~\cite{Azuma1979, Azuma1980}. In 1988, the only \( \beta \)-delayed four-neutron emitter so far, \( ^{17} \)B, was investigated~\cite{Dufour1988}. According to the Atomic Mass Evaluation in 2020 (AME2020)~\cite{Huang2021, Wang2021}, around \( 20 \) \% of the known nuclei are \( \beta \)-delayed neutron emitters. 

The accurate quantitative understanding of \( \beta \)-delayed neutron emission rates and spectra is necessary for several fields, including nuclear structure, nuclear astrophysics, and reactor applications~\cite{Kratz2023}. Detailed neutron spectroscopy is required to improve the knowledge of the decay schemes of neutron-rich isotopes far from the stability valley, and allows for more precise determinations of neutron emission probabilities (\( P_{xn} \))~\cite{Yokoyama2023}. Regarding nuclear technology in particular, precisely determined neutron emission spectra can improve the understanding of the kinematics and safety of new reactor concepts loaded with new types of fuels. This is achieved through applications such as reactor core characterization~\cite{Fiorina2012}, reactor shielding~\cite{Hajji2021}, and nuclear waste transmutation~\cite{Abanades2002}, among others.

During the last decades, the field of \( \beta \)-delayed neutrons has experienced an increased activity thanks to the advances in nuclear experimental techniques and radioactive ion beam (RIB) facilities producing increased yields of neutron-rich isotopes~\cite{Dimitriou2021}. Properties from individual precursors like the emission probability, \( \beta \)-feeding, and energy spectrum are being measured with advanced neutron detectors~\cite{Buta2000, Peters2016, Garcia2012}, digital data acquisition (DAQ) systems~\cite{Villamarin2023}, and high intensity RIBs~\cite{Spiller2006, Okuno2012, Moore2013, Catherall2017}.

\begin{table*}
  \caption{Relevant properties of the \( \beta \)-decays extracted from ENDF/B-VIII.0.\label{decays}}
  \begin{ruledtabular}
    \begin{tabular}{cccccc}
      Isotope       & \( T_{1/2} \) (s)     & \( Q_{\beta} \) (keV) & \( Q_{\beta n} \) (keV) & \( S_n \) (keV)      & \( P_{n} \) (\%)   \\\hline 
      \( ^{85} \)As & \( 2.021 \pm 0.010 \) & \( 9224.5 \pm 4.0 \)  & \( 4687.2 \pm 3.6 \)    & \( 4537.3 \pm 5.4 \) & \( 59.4 \pm 2.4 \) \\
      \( ^{86} \)As & \( 0.945 \pm 0.008 \) & \( 11541.0 \pm 4.3 \) & \( 5380.2 \pm 4.3 \)    & \( 6160.8 \pm 6.1 \) & \( 35.5 \pm 0.6 \) \\
    \end{tabular}
  \end{ruledtabular}
\end{table*}

In this paper, we report on the results obtained in the measurement of the \( ^{85, 86} \)As \( \beta \)-decays performed at the Ion Guide Isotope Separator On Line (IGISOL) facility~\cite{Moore2013} of the Accelerator Laboratory of the University of Jyväskylä. The \( ^{85, 86} \)As isotopes are fission products included in the priority list for reactor studies~\cite{Dillmann2014}, in part due to their significant contribution to the \( \beta \)-delayed neutron fraction in reactors, and the role of those in reactor control. For this reason, there has been an effort during the last years to improve the knowledge on the \( \beta \)-delayed neutron emission process of these isotopes~\cite{Agramunt2014, Garcia2020, Agramunt2023}, among others in the light-mass fission group. This work is a continuation of said effort, aiming to complete the available information by means of neutron spectroscopy with the MOdular Neutron time-of-flight SpectromeTER (MONSTER)~\cite{Garcia2012, Martinez2014}. Some relevant properties of the \( \beta \)-decays of \( ^{85, 86} \)As are presented in Table~\ref{decays}~\cite{Brown2018}. On the one hand, \( ^{85} \)As is a relatively well-known isotope~\cite{Franz1974, Kratz1979} in the vicinity of the \( N = 50 \) shell closure with a high production yield at IGISOL, which can be used as the experimental validation of the novel time-of-flight (TOF) analysis methodology developed for this work~\cite{RPC}. On the other hand, \( ^{86} \)As has a medium production yield at IGISOL and new spectroscopic information can be produced, shedding light on the nuclear structure of \( Z > 28 \), \( N > 50 \) nuclei.

\section{Experimental setup}\label{exp}

The \( ^{85, 86} \)As isotopes were produced by \( 25 \) MeV proton-induced fission reactions in a \( ^{nat} \)U target~\cite{Fadil2016, Pentilla2020}. The protons were accelerated with the MCC-30/15 cyclotron. In both cases, the isobars were separated from the bulk of fission products by the IGISOL dipole magnet with resolution \( M/\Delta M = 500 \)~\cite{Pentilla2020} and implanted on a movable tape inside of the \( \beta \)-detector. The data was collected in measurement cycles following the implantation-decay method. The implantation, decay, and activity removal periods were configured in a multi-timer module, which provides logic signals for synchronization, to maximize the activity of the neutron emitters with respect to the other implanted isobars. During the cycle, the ion beam was switched on and off by electrostatic deflection at the switchyard of the IGISOL beam line. The cycle also included an initial time interval for background measurement.

A picture of the whole experimental setup can be seen in Figure~\ref{setup}. The setup consists of a plastic scintillator for the detection of the emitted \( \beta \)-particles, MONSTER for the detection of the emitted neutrons, and two types of \( \gamma \)-ray detectors for the detection of the emitted \( \gamma \)-rays.

The emitted \( \beta \)-delayed neutrons were detected with MONSTER~\cite{Garcia2012, Martinez2014}, which is the result of an international collaboration between CIEMAT, JYFL, VECC, IFIC, and UPC within the framework of the DESPEC collaboration~\cite{Rubio2006, Mistry2022} of the NUSTAR facility~\cite{Nilsson2015} at FAIR~\cite{Spiller2006}. It was conceived for the measurement of the energy spectra of \( \beta \)-delayed neutrons with the TOF technique and the partial branching ratios to the excited states in the final nuclei by applying \( \beta\textrm{-}n\textrm{-}\gamma \) coincidences. The main characteristics of MONSTER are:

\begin{figure}
  \includegraphics[width=0.9\linewidth,trim={10cm 4cm 8cm 2cm},clip]{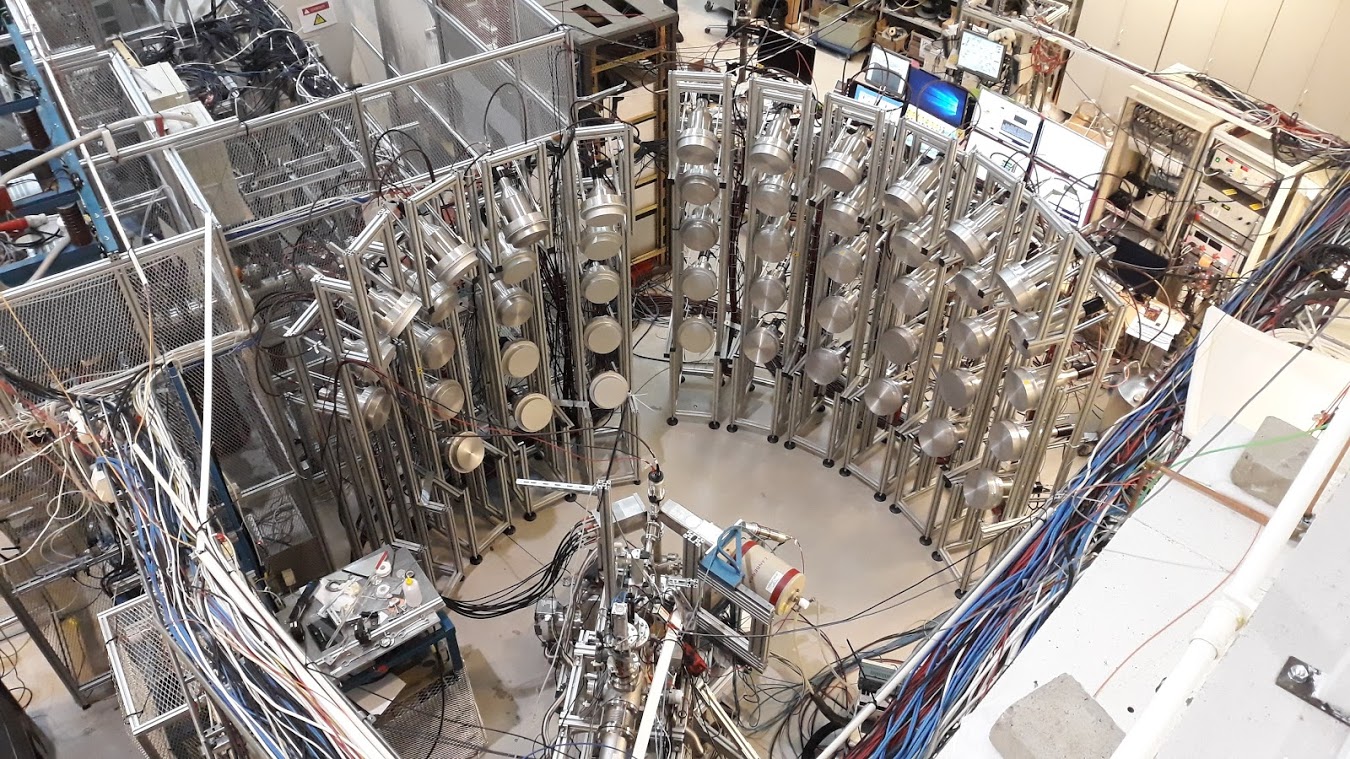}
  \caption{Experimental setup mounted at IGISOL.\label{setup}}
\end{figure}

\begin{itemize}
  \item low light output threshold (\( \sim 30 \) keVee);
  \item high intrinsic neutron detection efficiency (\( \sim 30 \) \%);
  \item neutron/\( \gamma \)-ray discrimination by their pulse shape;
  \item good time resolution (\( \sim 1 \) ns).
\end{itemize}

MONSTER is made of cylindrical cells of \( 200 \) mm diameter and \( 50 \) mm height, filled with either BC501A or EJ301 scintillating liquid. Each cell is coupled through a light guide of \( 31 \) mm thickness to either an R4144 or R11833 photomultiplier tube (PMT) from Hamamatsu. For this experiment, \( 48 \) cells were arranged in two different setups, with \( 30 \) and \( 18 \) cells at \( 2 \) and \( 1.5 \) m flight paths, respectively. The total solid angle covered by MONSTER was \( \sim 4 \) \%. The use of these conﬁgurations was decided to have a good neutron energy resolution while optimizing the available space and not compromising the total neutron detection efficiency. In addition, using different flight paths provided valuable information on the performance of the spectrometer in terms of energy resolution and signal-to-background ratio at different distances from the implantation point.

The \( \beta \)-detector consisted of a BC408 cup-like shaped plastic detector of \( 30 \) mm diameter, \( 50 \) mm length, and \( 2 \) mm thickness~\cite{Radivojevic2001}. The detector was wrapped with reflector foil to optimize the light collection. The design includes a cylindrical light guide of \( 30 \) mm diameter and \( 50 \) mm length for the coupling to an R5924-70 PMT from Hamamatsu. The \( \beta \)-detector was located at the end of the beam line, on top of the tape system and inside of the vacuum pipe, in such a way that the implantation in the tape could occur through a \( 10 \) mm aperture on the side of the detector. 

The signals from the \( \beta \)-detector and MONSTER were used as the start and stop signals, respectively, to measure the TOF of the neutrons.

Finally, an EXOGAM HPGe Clover detector~\cite{Azaiez1999} and four LaBr\( _3 \) crystals~\cite{Gramage2016} were used for the detection of the emitted \( \gamma \)-rays. However, due to low statistics, the data acquired with these detectors during the implantation runs was only used for monitoring the beam purity.

All the detector signals were registered with the Digital data AcquIsition SYstem (DAISY), a custom digital DAQ system developed at CIEMAT based on ADQ14DC digitizers from Teledyne SP Devices with \( 14 \) bits vertical resolution, \( 1 \) GHz sampling rate, and \( 4 \) channels~\cite{Villamarin2023}. The complete setup required \( 60 \) channels. The system integrates custom pulse shape analysis software also developed at CIEMAT to analyze the registered signals online. To analyze the registered frames, each frame is ﬁrst corrected for its baseline. After that, a smoothing ﬁlter is applied to reduce the noise. To determine the number of signals in the frame and their time, a digital constant fraction discriminator (CFD)
algorithm is applied. Once the CFD algorithm has been applied, the parameters of the detected signals are calculated with any of the several available pulse shape analysis routines, such as the digital charge integration method or fitting the detector's average pulse shape to the signals present in the frames~\cite{Guerrero2008}. The used pulse shape analysis routines did not add any dead time to the measurement, estimated to be less than \( 1 \) \%.

\section{Efficiency characterization}\label{eff}

An accurate characterization of the detection efficiency of the different detection systems used in the experiment is crucial to obtain a reliable result. In this section, the efficiency characterizations of the \( \beta \)-detector and of MONSTER are presented.

\subsection{\texorpdfstring{\( \beta \)}{Beta}-detector}

The \( \beta \)-detector efficiency was determined experimentally at two different endpoint energy values from the \( ^{92} \)Sr into \( ^{92} \)Y \( \beta \)-decay through the \( \beta\textrm{-}\gamma \) coincidence method using the HPGe and LaBr\( _{3} \) detectors. This decay has a half-life of \( 2.71 \pm 0.01 \) h, and the selected endpoint energy values were \( 565 \pm 9 \) and \( 1056 \pm 9 \) keV. The ions were implanted onto a non-moving tape for \( 1 \) h and the decay of the implanted ions was measured for \( 3 \) h.

Monte Carlo simulations performed with Geant4~\cite{Agostinelli2003} were used to determine the efficiency of the detector as a function of the endpoint energy. For this, several endpoint energies in the range of interest were simulated in order to obtain a smooth curve, which was then adjusted to the experimental values varying the deposited energy threshold. This procedure allowed to estimate the detection threshold to be of around \( 260 \) keV. Finally, all the efficiency curves compatible with the experimental values were selected in order to conclude that the experimental efficiency is reproduced with an uncertainty of \( 5 \) \%.

\begin{figure}
  \subfigure{\includegraphics[width=0.9\linewidth]{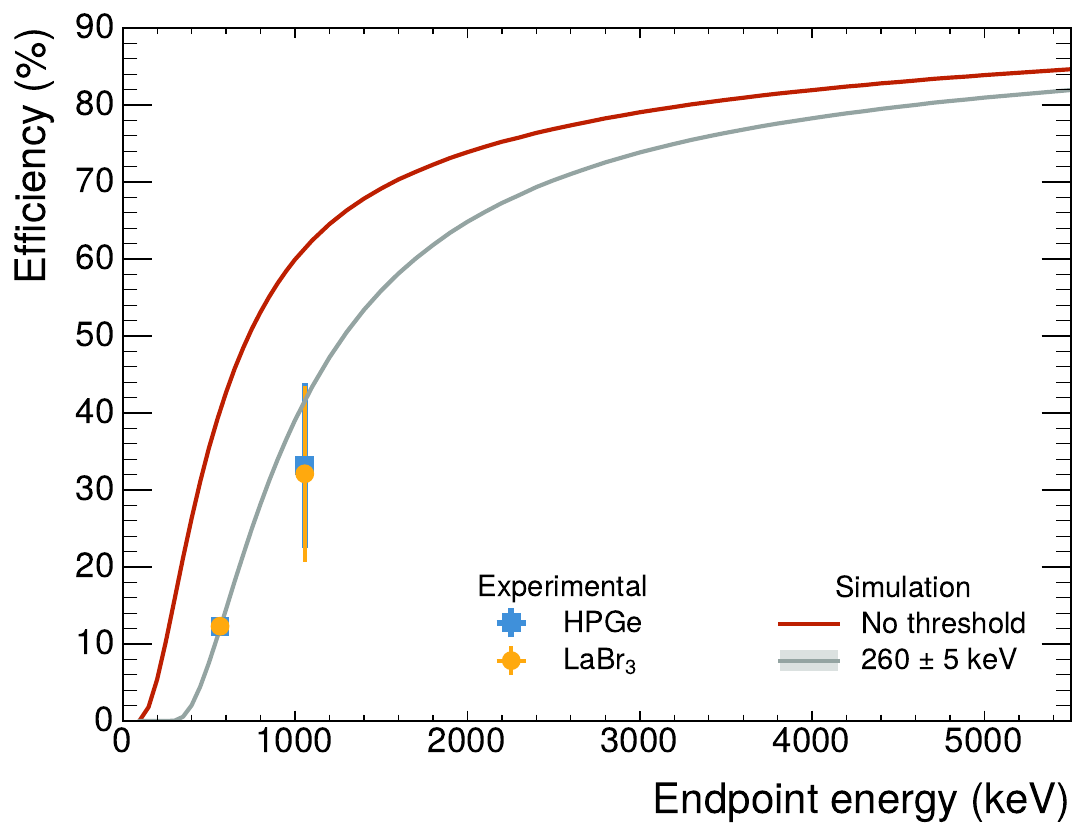}}
  \caption{\( \beta \)-efficiency determined from the \( ^{92} \)Sr into \( ^{92} \)Y \( \beta \)-decay and comparison with Monte Carlo simulations. The calculated efficiency curve with no threshold is shown for reference.\label{betaeff}}
\end{figure}

The results of the efficiency calculation and the comparison with the experimental data are shown in Figure~\ref{betaeff}.

\subsection{MONSTER}

\begin{figure}
  \includegraphics[width=0.9\linewidth]{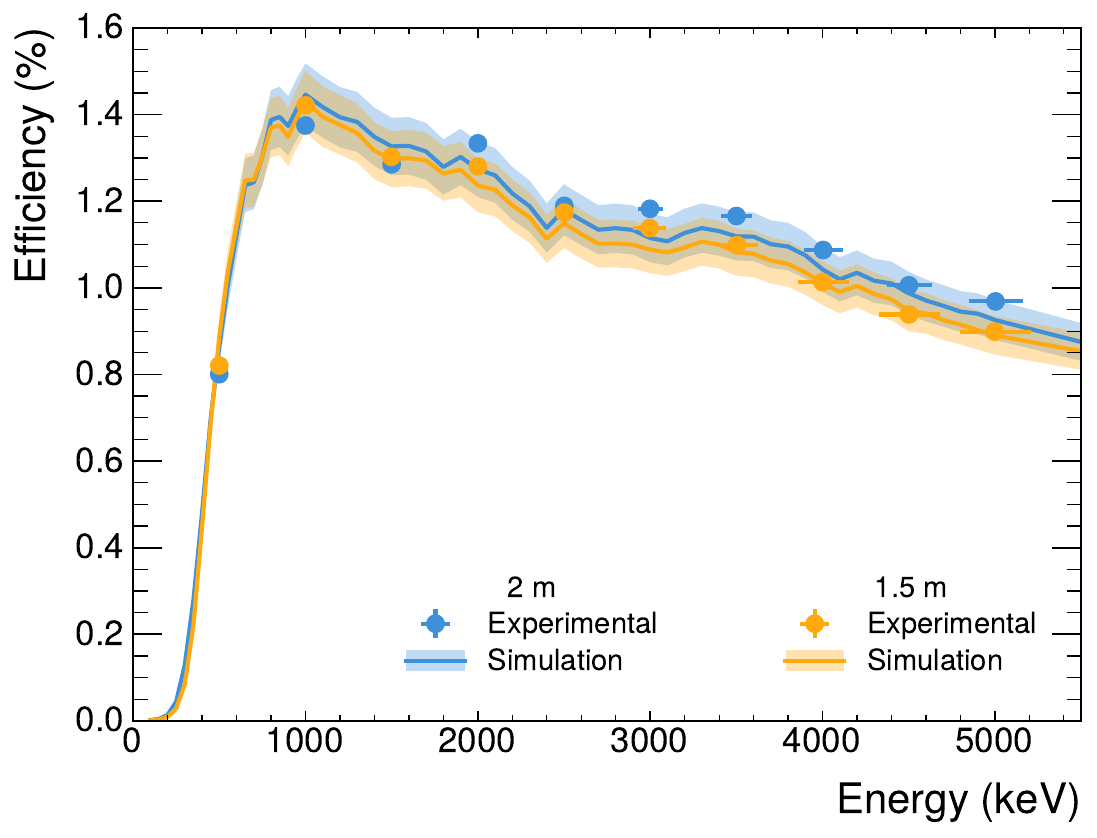}
  \caption{Total neutron detection efficiency for both MONSTER arrays determined from the \( ^{252} \)Cf source and compared with Monte Carlo simulations. The shaded regions correspond to the uncertainty in the simulations.\label{neff}}
\end{figure}

The MONSTER neutron detection efficiency was determined through Monte Carlo simulations and validated with the data from a TOF measurement with a \( 1.00 \pm 0.15 \) GBq \( ^{252} \)Cf source.

The Monte Carlo simulations were performed with the light output response obtained from previous calibrations with quasi-monochromatic neutron beams~\cite{Martinez2014}. To obtain a smooth curve, several energies in the range of interest were simulated. The resolution and threshold determined experimentally were also taken into account.

To obtain the experimental values of the detection efficiency, the \( \gamma \)-rays from the spontaneous fission of \( ^{252}\)Cf detected in the LaBr\(_{3}\) detectors served for establishing a TOF between the emission and the detection of the neutrons. Neutron signals in MONSTER were selected by performing a pulse shape discrimination (PSD) analysis based on the digital charge integration method. Selecting the neutron signals in this way allowed for a reduction of the background due to \( \gamma \)-rays of one order of magnitude. The response of MONSTER to different neutron energies was obtained by setting different cuts in the TOF spectrum. These responses were compared to the ones obtained from a simulation of the neutron emission of the \( ^{252}\)Cf source using the same TOF cuts. The simulation was normalized to the measured TOF spectrum. This comparison allowed to obtain the absolute value of the efficiency for every energy.

The results of the efficiency characterization of MONSTER are shown in Figure~\ref{neff}. As can be seen in the figure, there is an excellent agreement, within an uncertainty of 5 \%, between the experimental values and the Monte Carlo simulations of the MONSTER arrays for the two different flight paths.

\section{Data analysis and results}\label{res}

In general terms, the analysis methodology applied for this work consists of:

\begin{itemize}
  \item Fitting the solutions of the Bateman equations to the growth and decay curves followed by the implanted ions, for obtaining both the total number of ions being implanted and the total number of decays. This information was obtained from the \( \beta \)-activity registered with the \( \beta \)-detector.
  \item Unfolding the \( \beta \)-delayed neutron TOF spectra resulting from the neutron events detected with MONSTER correlated with \( \beta \)-particles detected with the \( \beta \)-detector using a novel data analysis methodology for producing the neutron energy distributions and obtaining the total number of emitted neutrons.
  \item Obtaining the neutron emission probability per decay, the \( P_n \) value, from the ratio of the total number of emitted neutrons to the total number of decays.
\end{itemize}

A suite of C++ programs integrating ROOT~\cite{Brun1997} has been developed for the data analysis presented in this work. Besides, the figures showing the results have been produced using some of the Python packages provided by the Scikit-HEP project~\cite{Rodrigues2020}. A full description of this analysis can be found in Reference~\cite{Thesis}.

\subsection{\texorpdfstring{\( \beta \)}{Beta}-activity distributions}\label{beta}

The time evolution of the \( \beta \)-activity during each measurement cycle due to all isotopes in the decay chain is governed by the Bateman equations. The experimental \( \beta \)-activity distribution is fitted with a function that describes the contribution of each of the isotopes involved in the decay chain:

\begin{equation}
  A(t) = \sum_{i=1}^{n} \overline{\epsilon_{i}}\lambda_{i}N_{i}(t),
  \label{BetaActivity}
\end{equation}

where \( \overline{\epsilon_{i}} \) is the average efficiency to detect a \( \beta \)-particle coming from the \( \beta \)-decay of the \( i \)-th member of the chain, \( \lambda_{i} \) is the decay constant of the \( i \)-th member of the chain, and \( N_{i}(t) \) is the number of nuclei of the \( i \)-th member of the chain given by the solution of the Bateman equations~\cite{Skrable1974}. This solution depends on the respective implantation rate of each isotope \( R_{i} \), which is the magnitude that we will obtain from the fit.

\begin{figure}
  \subfigure[\( A = 85 \)]{\includegraphics[width=0.9\linewidth]{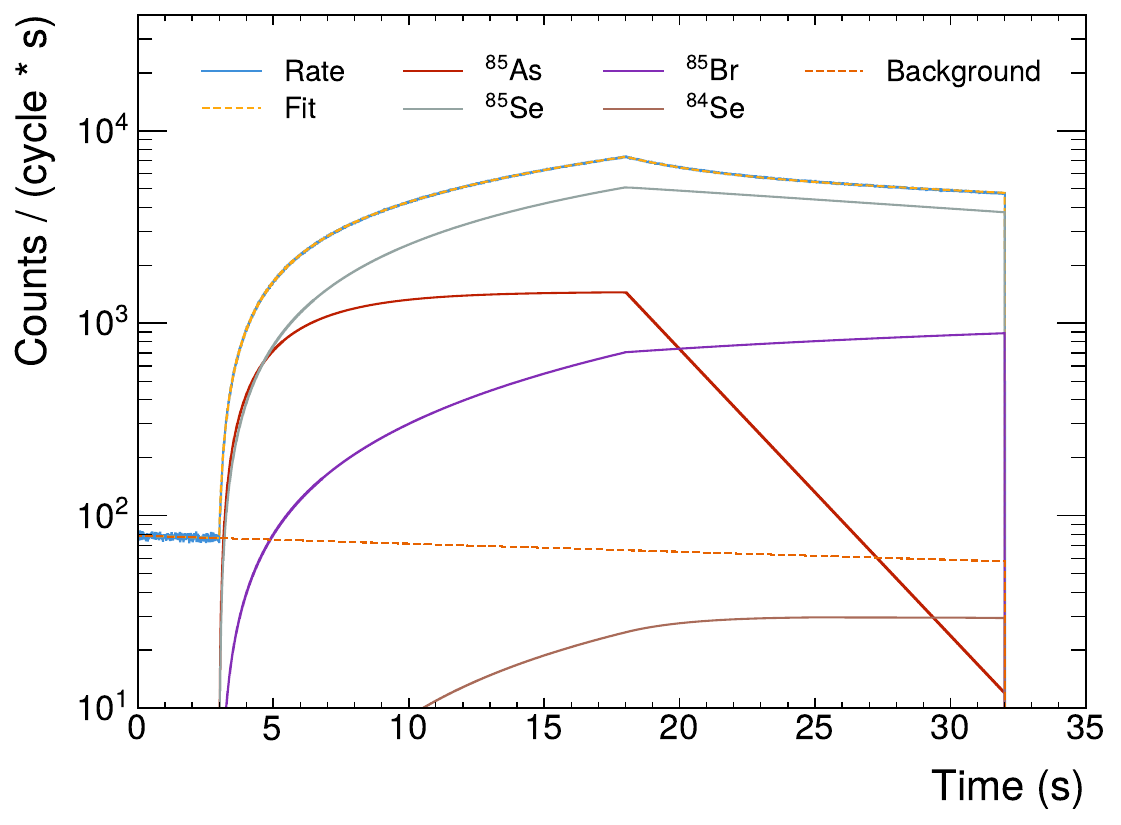}}
  \subfigure[\( A = 86 \)]{\includegraphics[width=0.9\linewidth]{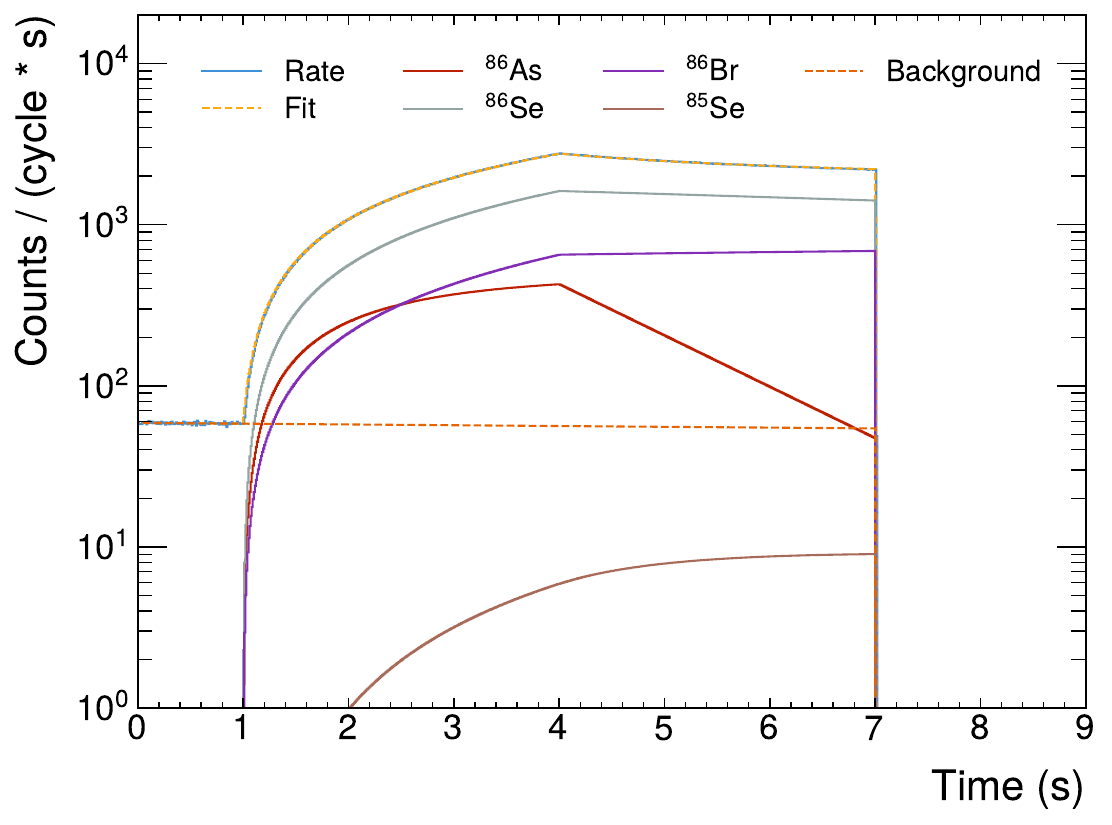}}
  \caption{Time evolution of the \( \beta \)-activity distributions and the corresponding fits, including individual contributions, for both cases.\label{betafit}}
\end{figure}

The experimental \( \beta \)-activity distributions for both decays, together with the total fits and the individual contributions, can be seen in Figure~\ref{betafit}. As can be seen in the figure, the duration of the measurement cycle in the case of \( ^{86} \)As was shorter in order to maximize its contribution with respect to the other implanted isobars.

The results of the fits for both decays are presented in Table~\ref{betafitresults}. For both decays, all the isobars with a half-life comparable to the duration of the measurement cycle were taken into account. In the table are also presented the calculated average detection efficiencies as well as the total number of decays of each of the isotopes. The average detection efficiencies for each of the \( \beta \)-decays were calculated through Monte Carlo simulations using the data available in ENDF/B-VIII.0~\cite{Brown2018}.

\begin{table}
  \caption{Results of the fit to the \( \beta \)-activity curve along with the calculated average efficiencies and the total number of decays.\label{betafitresults}}
  \begin{ruledtabular}
    \begin{tabular}{cccc}
      \multirow{2}{*}{Isotope} & \multirow{2}{*}{\( \overline{\epsilon} \) (\%)} & \multirow{2}{*}{\( R \) (ions/s)} & Total decays          \\
                               &                                                 &                                   & (\( \times 10^{6} \)) \\\hline 
      \( ^{85} \)As            & \( 80.8 \pm 4.0 \)                              & \( 1800 \pm 100 \)                & \( 48 \pm 3 \)        \\
      \( ^{85} \)Se            & \( 76.0 \pm 3.8 \)                              & \( 23300 \pm 900 \)               & \( 239 \pm 9 \)       \\
      \( ^{85} \)Br            & \( 69.7 \pm 3.5 \)                              & \( 13900 \pm 2500 \)              & \( 42 \pm 5 \)        \\
      \( ^{84} \)Se            & \( 55.1 \pm 2.8 \)                              & \( 0 \pm 0 \)                     & \( 1.8 \pm 0.1  \)    \\\hline 
      \( ^{86} \)As            & \( 83.5 \pm 4.2 \)                              & \( 570 \pm 40 \)                  & \( 12.7 \pm 0.8 \)    \\
      \( ^{86} \)Se            & \( 72.9 \pm 3.6 \)                              & \( 16100 \pm 400 \)               & \( 74 \pm 2 \)        \\
      \( ^{86} \)Br            & \( 77.5 \pm 3.9 \)                              & \( 21500 \pm 2100 \)              & \( 30 \pm 3 \)        \\
      \( ^{85} \)Se            & \( 76.0 \pm 3.8 \)                              & \( 0 \pm 0 \)                     & \( 0.32 \pm 0.02  \)  \\
    \end{tabular}
  \end{ruledtabular}
\end{table}

\subsection{Neutron energy distributions}\label{neutron}

The energy distributions of the neutrons emitted after the \( \beta \)-decays of \( ^{85, 86} \)As were obtained from the measured TOF spectra with an innovative methodology~\cite{RPC}, based on the iterative Bayesian unfolding method~\cite{DAgostini1995} and precise Monte Carlo simulations. Due to the complex nature of neutron interactions, a good physics model and correct implementation of the detector geometries and surrounding materials are very important for calculating accurate TOF distributions. For this reason, the G4ParticleHP model~\cite{Mendoza2011} included in Geant4 was used, since it has been validated extensively for the simulation of MONSTER modules~\cite{Martinez2014, Garcia2011, Garcia2017}. This high-precision model includes the necessary models and data for a complete description of neutron-induced alpha production reactions on carbon. The detailed simulations performed in this work include the parametrized light output functions of the scintillators for electrons, protons, and heavier ions as a function of their energy. In addition, the experimental setup was implemented with a high level of detail, including all the MONSTER modules, ancillary detectors surrounding the implantation point, and other elements like the structures holding the detectors and the floor~\cite{RPC}.

\subsubsection{\texorpdfstring{\( ^{85} \)}{85}As}\label{85as}

\begin{figure}
  \subfigure[Light versus TOF]{\includegraphics[width=0.9\linewidth]{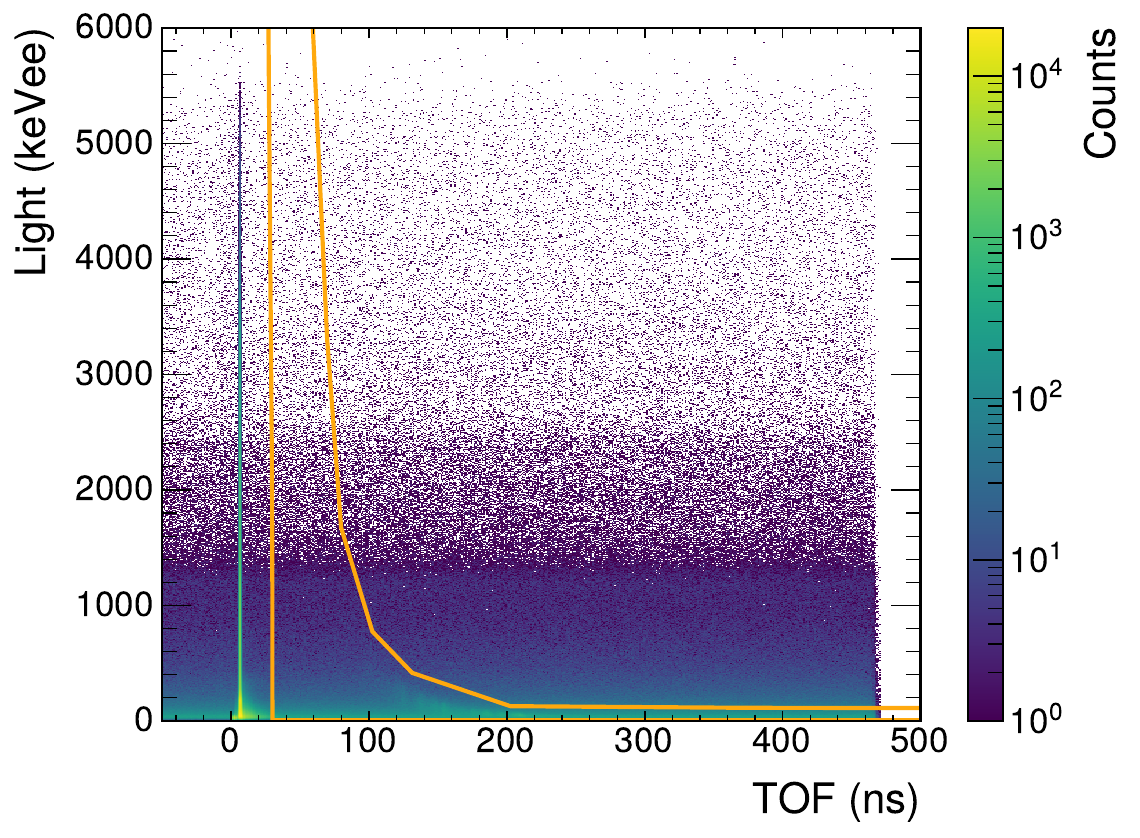}}
  \subfigure[\( PSD \) versus TOF]{\includegraphics[width=0.9\linewidth]{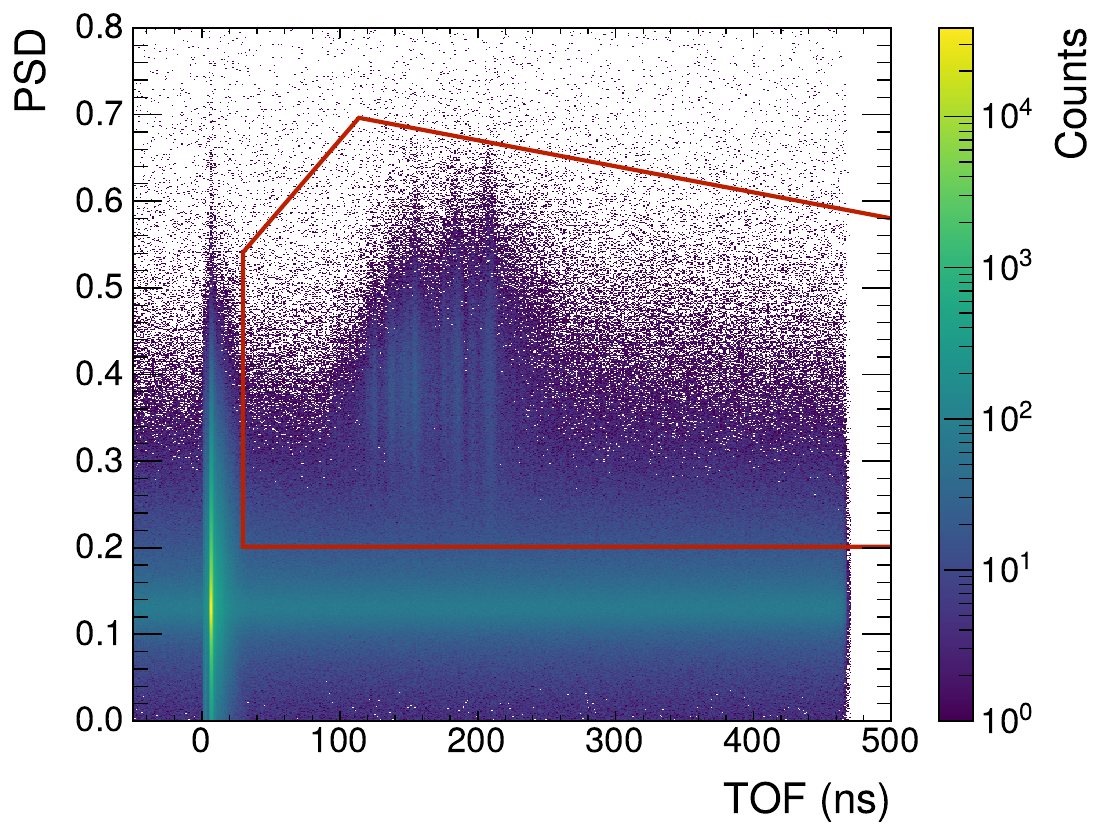}}
  \subfigure[\( PSD \) versus light]{\includegraphics[width=0.9\linewidth]{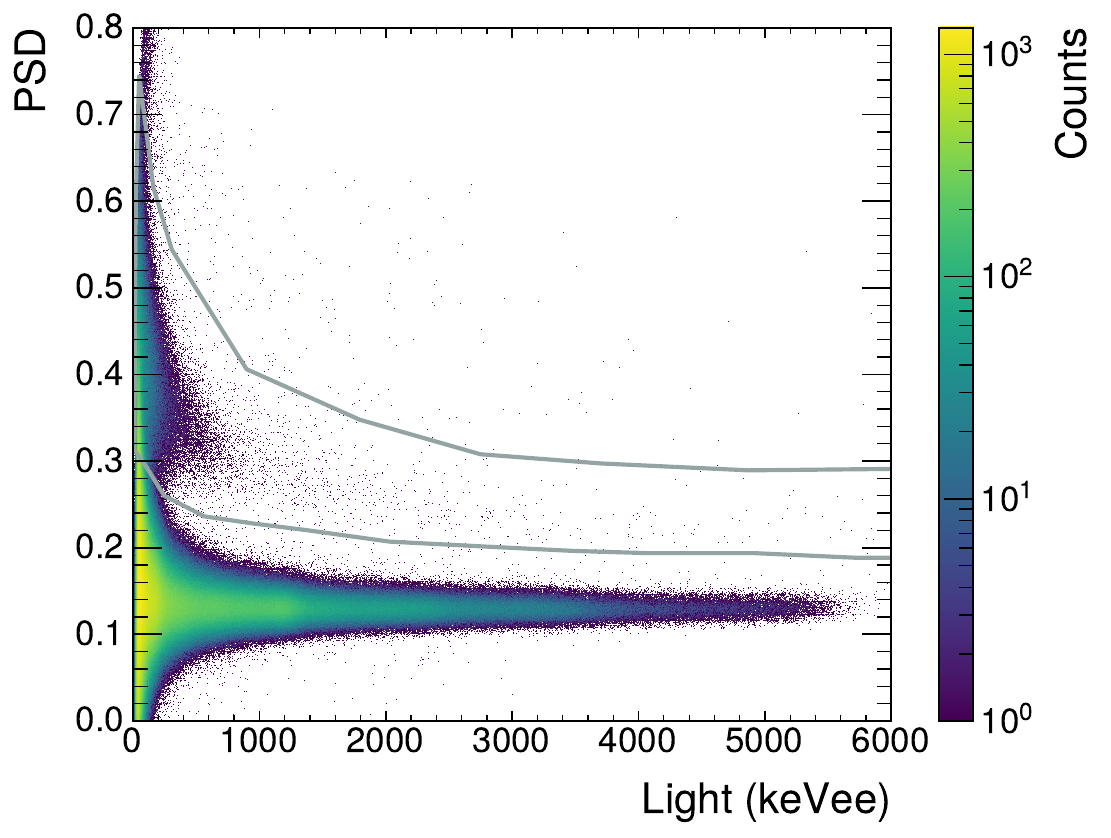}}
  \caption{Neutron selection cuts studied.\label{neutroncuts}}
\end{figure}

In principle, the measured TOF spectra contain the contribution from both \( \gamma \)-rays and neutrons, but the contribution due to neutrons can be separated by applying a cut in the \( PSD \) versus light spectrum. However, other neutron selection cuts, such as the typical one used in TOF measurements with plastic scintillators, were investigated as well. All the neutron selection studied in this work are shown in Figure~\ref{neutroncuts} for the MONSTER array at \( 2 \) m, in the case of the \( ^{85} \)As \( \beta \)-decay.

\begin{figure}
  \subfigure{\includegraphics[width=0.9\linewidth]{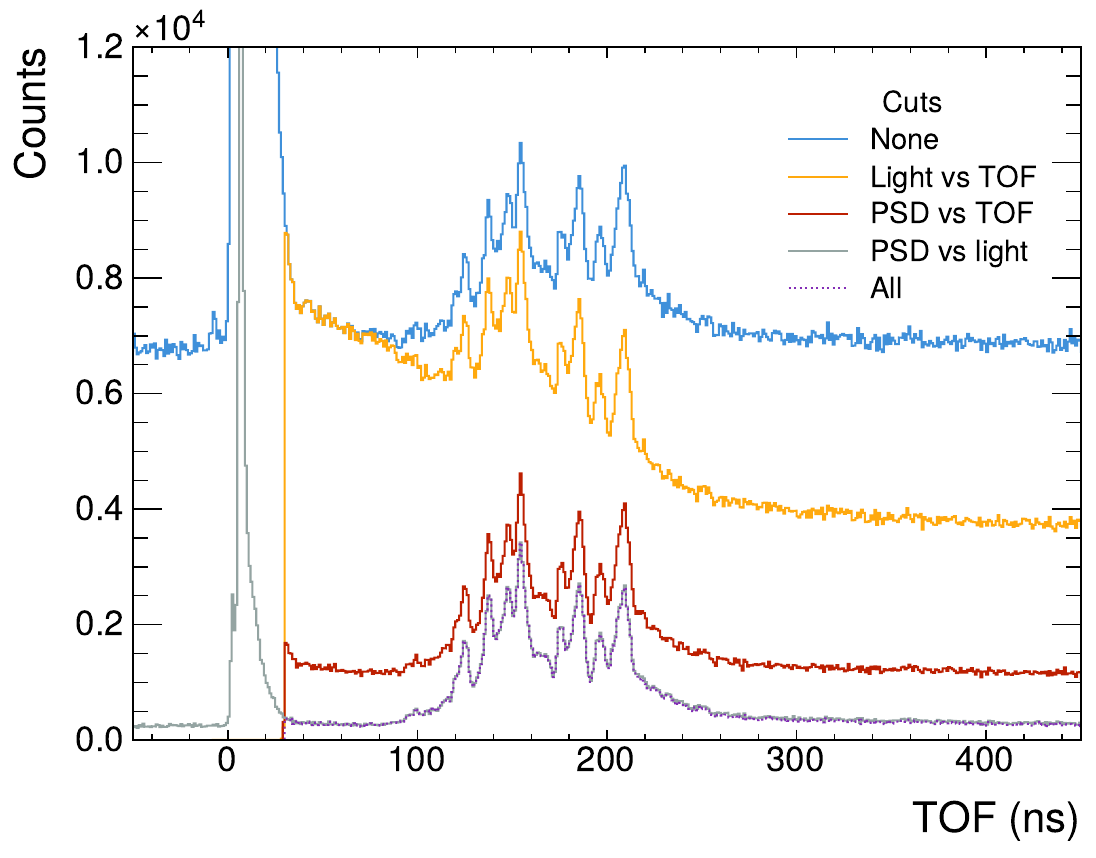}}
  \caption{\( ^{85} \)As measured TOF spectra with the MONSTER array at \( 2 \) m as registered and with several neutron selection cuts.\label{tof}}
\end{figure}

The raw TOF spectrum and the TOF spectra resulting from the application of each of the neutron selection cuts, separately and all of them combined, can be seen in Figure~\ref{tof}. From the figure, it is clear that the cut in the \( PSD \) versus light spectrum is the most effective. This cut provides the best separation, which is almost as good as when applying all cuts combined. Thus, it was decided to only use the cut in the \( PSD \) versus light spectrum. With this neutron selection cut, the uncorrelated \( \gamma \)-ray background is reduced by more than one order of magnitude.

\begin{figure}
  \subfigure{\includegraphics[width=0.9\linewidth]{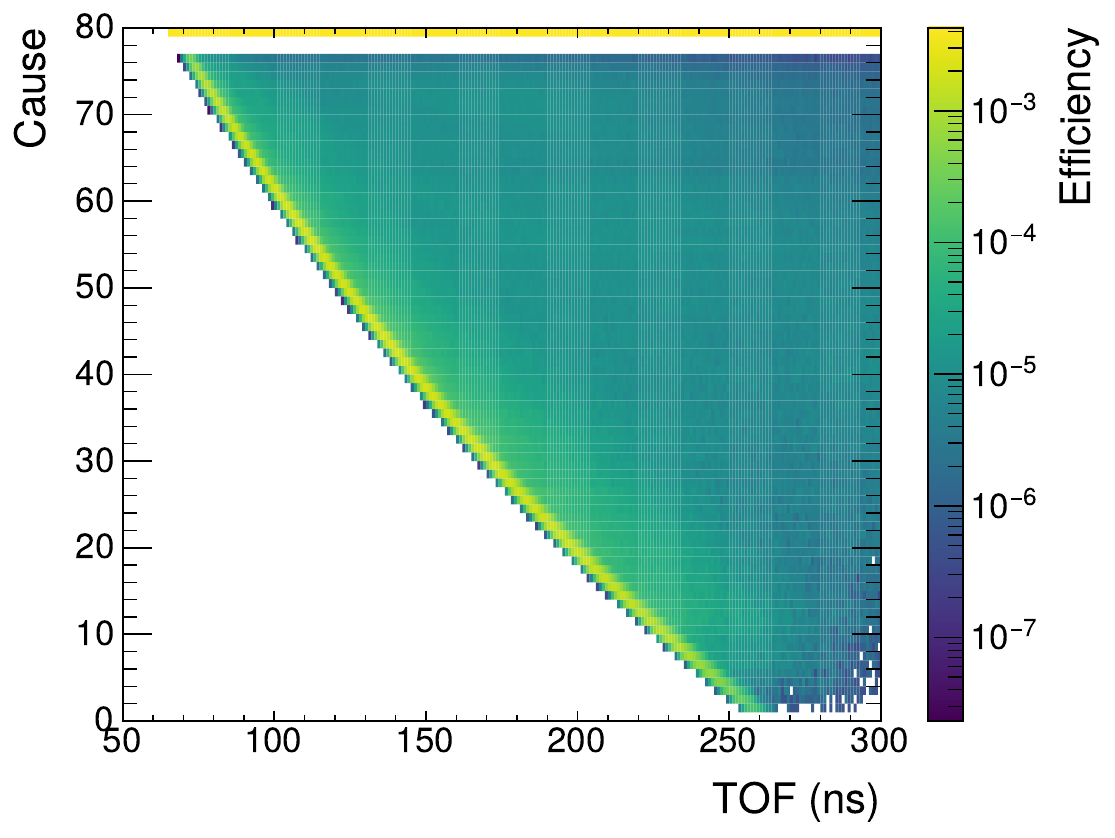}}
  \caption{Response matrix used for the unfolding of the \( ^{85} \)As TOF spectrum for the MONSTER array at \( 2 \) m. Each cause but the last one corresponds to the response of a different neutron energy interval. The last one is a flat response to account for the uncorrelated background.\label{responsematrix}}
\end{figure}

In order to perform the unfolding, the construction of the response matrix is crucial to achieve an accurate result. The response matrices used for this analysis covered the whole energy range up to the \( Q_{\beta n} \) values of the respective decays, in energy intervals chosen according to the energy resolution of the detection system. They included many experimental effects, such as the exact flight path of each detector, the TOF resolution of the system, the light production and detection threshold of each detector, and also included an extra flat contribution to account for the uncorrelated \( \gamma \)-ray background. As an example, the response matrix that has been used for the MONSTER array at \( 2 \) m in the case of the \( ^{85} \)As \( \beta \)-decay can be seen in Figure~\ref{responsematrix}.

\begin{figure}
  \subfigure{\includegraphics[width=0.9\linewidth]{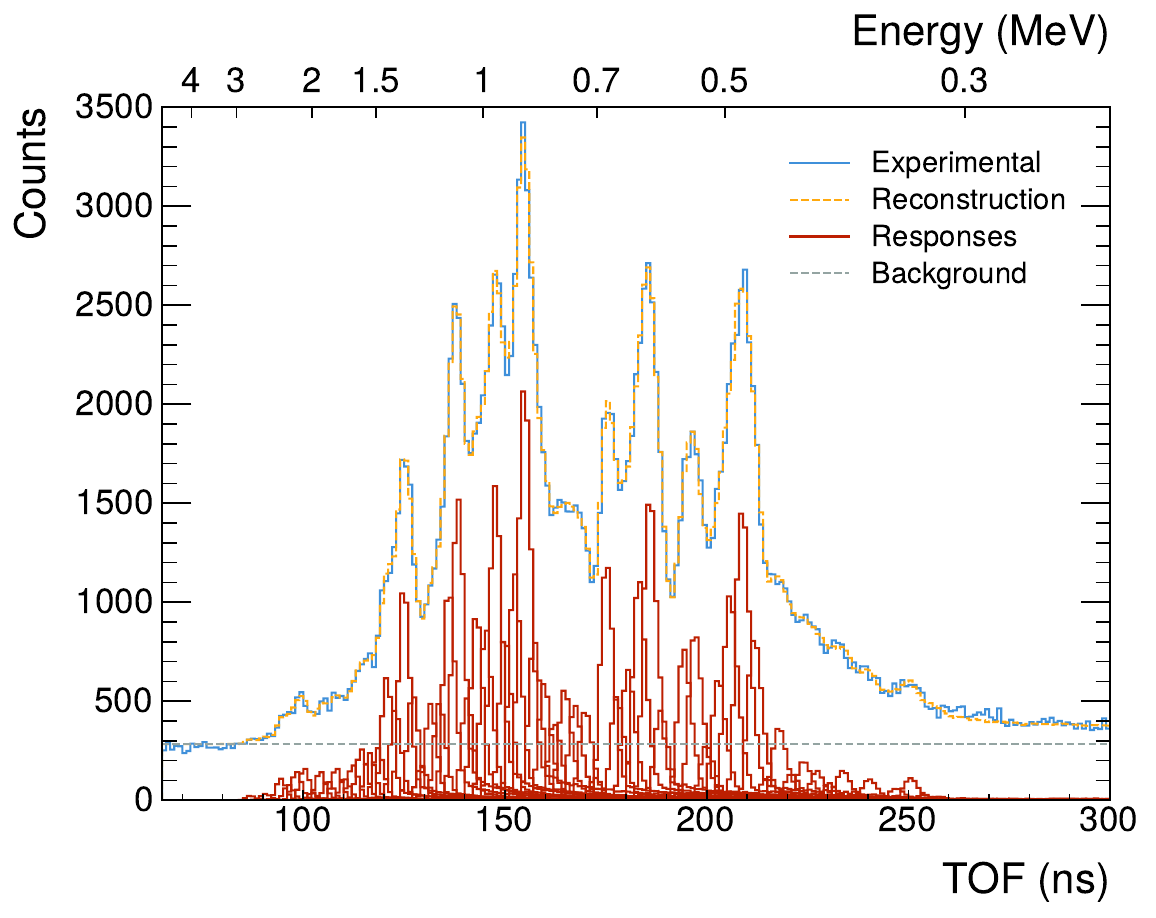}}
  \caption{Experimental TOF spectrum obtained with the MONSTER array at \( 2 \) m for the \( ^{85} \)As decay. The reconstructions, together with the contributions of the individual responses and the uncorrelated background, are also shown.\label{rec_as85}}
\end{figure}

The experimental TOF spectrum measured with the MONSTER array at \( 2 \) m in the case of the \( ^{85} \)As \( \beta \)-decay is shown in Figure~\ref{rec_as85}. In the figure are also shown the reconstruction of the TOF spectrum and the contributions of the individual responses and of the uncorrelated background.

\begin{figure*}
 \subfigure{\includegraphics[width=0.8\linewidth]{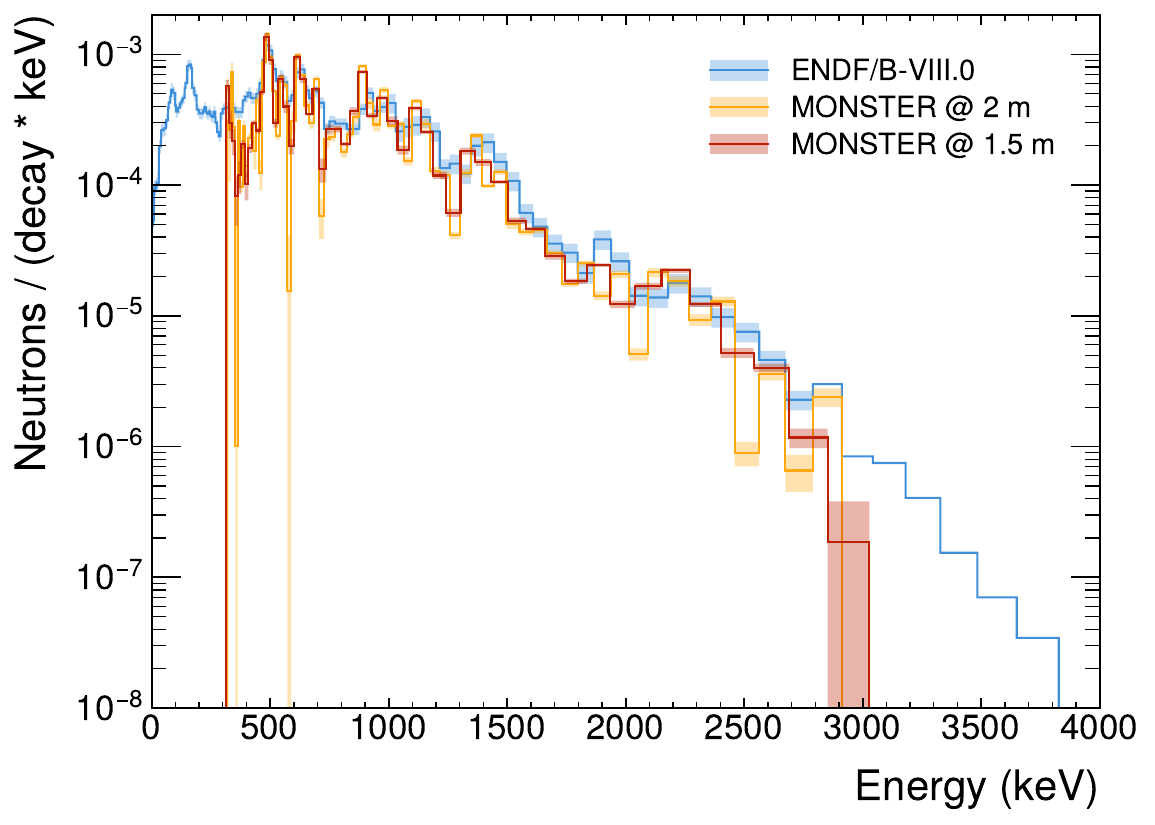}}
 \caption{Neutron energy distributions obtained with both MONSTER arrays for the \( ^{85} \)As decay, compared to evaluated data.\label{ned_as85}}
\end{figure*}

The neutron energy distributions following the \( \beta \)-decay of \( ^{85} \)As obtained with both MONSTER arrays are presented in Figure~\ref{ned_as85}. The agreement of the results obtained with both arrays is excellent within their energy resolution. The results include an estimation of both statistical and systematic uncertainties. On the one hand, the statistical uncertainties have been directly obtained from the unfolding's covariance matrix. On the other hand, the systematic uncertainties have been obtained by performing different unfoldings. First, several unfoldings using different response matrices were performed. These response matrices were obtained by varying the most relevant parameters of the measurement---flight path, TOF resolution, and efficiency of the different detectors---within their respective uncertainties. Second, to quantify the uncertainty due to the Bayesian unfolding method itself, an unfolding using the Maximum Entropy method described in Reference~\cite{Tain2007} was performed. The spectra obtained with all the different unfoldings were compared to the results, and the differences were taken as the systematic uncertainties. The total uncertainties are the sum of the statistical and systematic uncertainties.

In the figure, the obtained results are also compared to existing data~\cite{Brady1989, Brown2018}. The existing data has been rebinned to the binning of the results of this work. The results obtained for \( ^{85} \)As are in excellent agreement with evaluations, which include experimental data obtained by the Kernchemie Mainz group~\cite{Franz1974, Kratz1979} and are supplemented with theoretical calculations~\cite{Kawano2008}. The differences observed at low energies, below \( 300 \) keV, are due to the neutron detection threshold of MONSTER\@. At high energies, above \( 2800 \) keV, where the evaluated data is based on theoretical calculations, the predicted intensity is not observed. This could be due to an overestimation of the uncorrelated \( \gamma \)-ray background, which is a result of the unfolding. However, the systematic effects study did not show significant variations in the flat background contribution, and given that such neutron emission is not seen in any of the MONSTER arrays, with very different signal-to-background ratios, it seems that no neutrons at such high energies are actually emitted. Finally, it is worth noting that the uncertainties of the results obtained in this work are smaller than those reported in the literature.

\begin{figure}
  \subfigure{\includegraphics[width=0.9\linewidth]{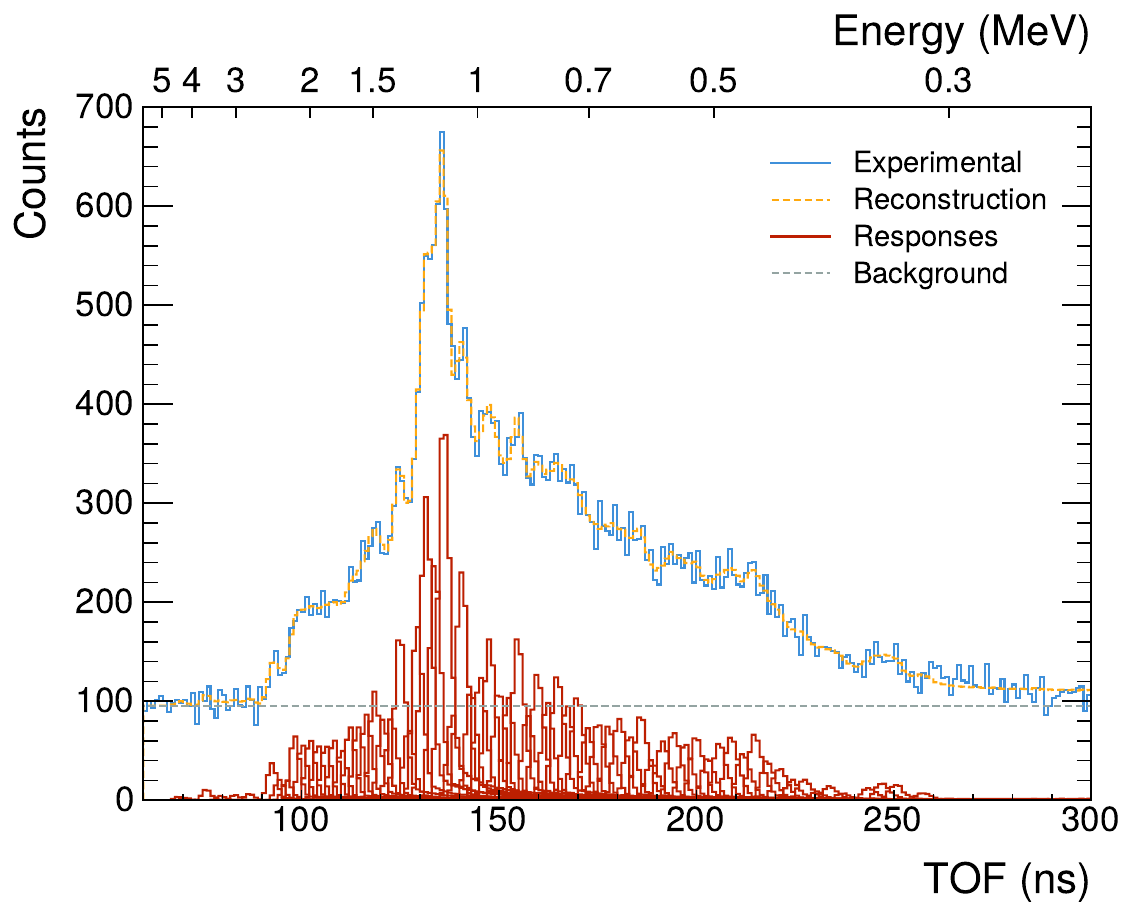}} 
  \caption{Experimental TOF spectrum obtained with the MONSTER array at \( 2 \) m for the \( ^{86} \)As decay. The reconstructions, together with the contributions of the individual responses and the uncorrelated background, are also shown.\label{rec_as86}}
\end{figure}

\begin{figure*}
 \subfigure{\includegraphics[width=0.8\linewidth]{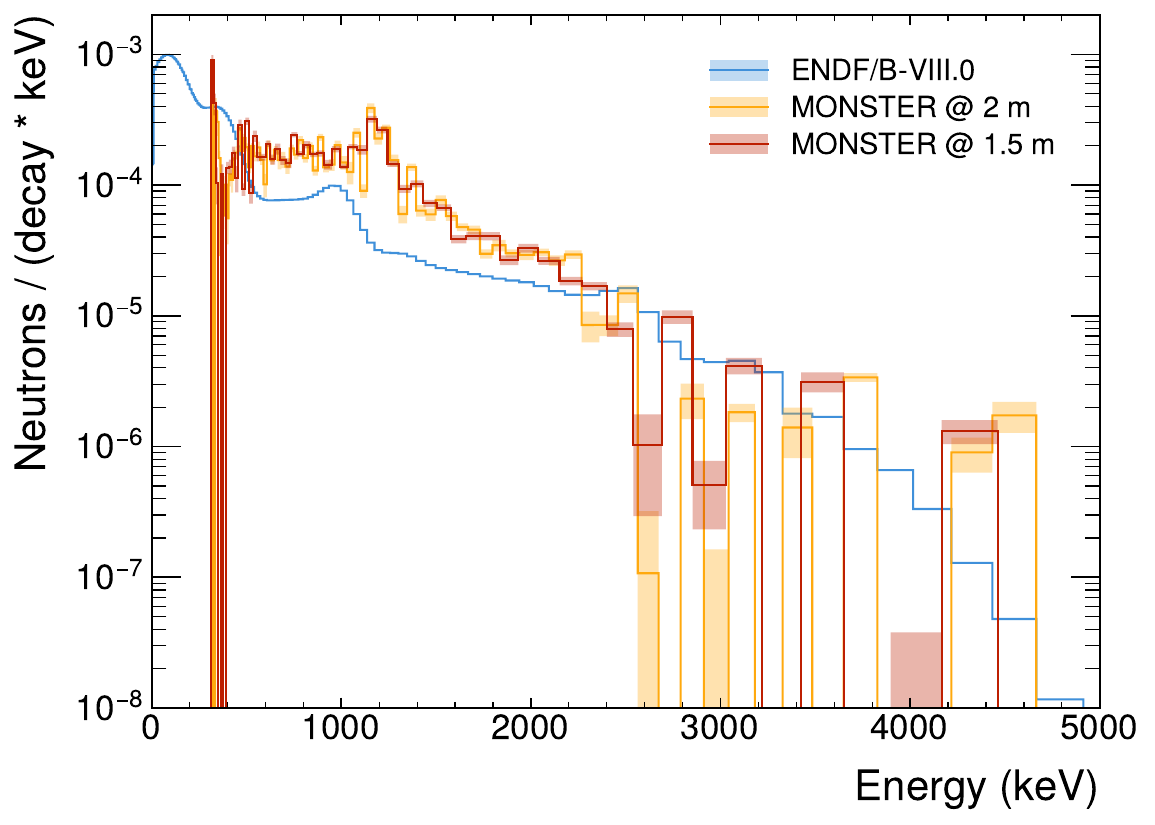}} 
 \caption{Neutron energy distributions obtained with both MONSTER arrays for the \( ^{86} \)As decay, compared to evaluated data.\label{ned_as86}}
\end{figure*}

The total number of neutrons emitted in the \( \beta \)-decay of \( ^{85} \)As is a direct product of the unfolding and, together with the total number of \( ^{85} \)As \( \beta \)-decays obtained in Section~\ref{beta}, yields an estimated \( P_{n} \) value of \( 40 \pm 4 \) \% in the case of the MONSTER array at \( 2 \) m and of \( 41 \pm 4 \) \% in the case of the MONSTER array at \( 1.5 \) m. These results are once again perfectly compatible between them within their uncertainties. The average \( P_{n} \) value obtained is \( 40.5 \pm 2.8 \) \%. It is worth noting that this result is biased, since it does not account for the undetected neutrons due to the different detection thresholds of the detectors. Indeed, the \( P_{n} \) value obtained in this work is below the ones reported in the literature, either the \( 59.4 \pm 2.4 \) \% from evaluations~\cite{Brown2018, Liang2020} or the \( 63.1 \pm 1.0 \) \% from more precise recent measurements with \( ^{3} \)He proportional counters~\cite{Agramunt2014, Agramunt2023}.

\subsubsection{\texorpdfstring{\( ^{86} \)}{86}As}\label{86as}

In the case of the \( ^{86} \)As \( \beta \)-decay, the TOF spectrum measured with the MONSTER array at \( 2 \) m is presented in Figure~\ref{rec_as86}. In the figure are also shown the reconstruction of the TOF spectrum and the contributions of the individual responses and of the uncorrelated background.

The neutron energy distributions following the \( \beta \)-decay of \( ^{86} \)As obtained with both MONSTER arrays are presented in Figure~\ref{ned_as86}. As can be seen in the figure, the agreement of the results obtained with both arrays is excellent within their energy resolution. These results also include an estimation of both statistical and systematic uncertainties, obtained in the same way as previously explained. No previous experimental energy distribution has been found for \( ^{86} \)As in the literature. As such, the evaluated data includes only theoretical calculations. There is some degree of agreement between the results of this work and the evaluated data, but there are also some differences worth noting. The increased intensity region predicted at around \( 1000 \) keV is observed between \( 1200 \) and \( 1300 \) keV. The intensity at high neutron energies, above \( 3000 \) keV, is observed in this case, and could even be larger than expected. In general, the measured neutron energy distribution seems more intense at higher neutron energies than the calculated one, which predicts a more intense emission of lower-energy neutrons. The results obtained in this work will be key for improving both the evaluations of the \( ^{86} \)As \( \beta \)-decay and the theoretical models.

The \( P_{n} \) value of the \( ^{86} \)As \( \beta \)-decay has also been estimated, yielding \( 23 \pm 2 \) \% in the case of the MONSTER array at \( 2 \) m and \( 24 \pm 2 \) \% in the case of the MONSTER array at \( 1.5 \) m. The average \( P_{n} \) value obtained is \( 23.5 \pm 1.4 \) \%. As in the case of the \( ^{85} \)As \( \beta \)-decay, this value is below the \( 35.5 \pm 0.6 \) \% reported in the literature~\cite{Brown2018, Liang2020, Agramunt2014, Garcia2020, Agramunt2023}, which is once again attributed to the detection thresholds of the detectors.

\section{Summary and conclusions}\label{conc}

This work is a successful commissioning and measurement with MONSTER, a neutron TOF spectrometer with excellent neutron/\( \gamma \)-ray discrimination capabilities and neutron energy resolution, and its DAQ system DAISY. 

The efficiency of the different detectors has been obtained experimentally and reproduced with accurate Monte Carlo simulations.

As part of this work, the innovative analysis methodology developed to obtain the neutron energy distributions from TOF measurements has been experimentally validated. This methodology, combined with the fit of the \( \beta \)-activity curves with the Bateman equations, allows to estimate the \( P_n \) value of the \( \beta \)-decays.

The \( \beta \)-delayed neutron energy spectrum of \( ^{85} \)As has been procured. This result is in excellent agreement with previous experimental data and evaluations, and serves as the experimental validation of the developed analysis methodology.

Finally, the \( \beta \)-delayed neutron energy spectrum of \( ^{86} \)As has been procured for the first time. The obtained energy distribution shares some similarities with the theoretical calculations that have been included in evaluations, but it also presents many differences. In particular, a stronger neutron intensity at higher energies than previously predicted has been observed. The results obtained in this work will be key to improve both the evaluations of the \( ^{86} \)As \( \beta \)-decay and the theoretical models.

The results presented in this work will be delivered to the EXFOR library~\cite{Otuka2014}.

\begin{acknowledgments}
  This work was partially supported by the I+D+i grants FPA2016-76765-P, PGC2018-096717-B-C21, and PID2022-142589OB-I00 funded by MCIN/AEI/10.13039/501100011033, and by the European Commission H2020 Framework Programme project SANDA (Grant agreement ID:\ 847552). 

  This project has received funding from the European Union’s Horizon 2020 research and innovation programme under grant agreements No. 771036 (ERC CoG MAIDEN) and the Research Council of Finland projects No. 275389, 284516, 312544, and 354968.

  This work was partially funded by Spanish MCIN/AEI/10.13039/501100011033 under grants RTI2018-098868-B-I00 and PID2021-126998OB-I0. J. B. acknowledges support from the UCM Spanish MIU and EU Next-Generation funds under grants No. CT27/16-CT28/16 and CT31/21. 
\end{acknowledgments}

\bibliography{bibliography}

\end{document}